\documentclass[twocolumn]{aa}
\usepackage{epsfig}
\usepackage{txfonts}
\begin{document}
   \title{An improved deprojection and PSF-deconvolution technique for
   galaxy-cluster X-ray surface-brightness profiles}

   \author{J.H. Croston
          \inst{1,2}
          \and
          M. Arnaud\inst{2}
	  \and
	  E. Pointecouteau\inst{3}
	  \and
	  G.W. Pratt\inst{4}
}

   \offprints{J.H. Croston}

   \institute{Centre for Astrophysics Research, Science and Technology
              Research Institute, University of Hertfordshire,
              Hatfield, AL10 3LB, UK\\
              \email{jcroston@star.herts.ac.uk}
         \and
	 CEA Saclay, Service d'Astrophysique, Orme des Merisiers, Bat 709, Gif-sur-Yvette Cedex, 91191, France\\
	     \and
CESR, 9 av. du colonel Roche, BP4346, 31028 Toulouse Cedex 4, France\\
\and
MPE Garching, Giessenbachstrasse, 85748 Garching, Germany\\}

   \date{}

   \abstract{ We have developed a regularisation procedure for the
     direct deprojection and PSF-deconvolution of X-ray surface
     brightness profiles of clusters of galaxies. This procedure
     allows us to obtain accurate density profiles in a
     straightforward manner from X-ray observations (in particular
     data from {\it XMM-Newton}, where the PSF correction is
     important), while retaining information about substructure in the
     gas distribution, in contrast to analytic modelling of the
     profiles. In addition to describing our procedure, we present
     here a detailed investigation of the accuracy of the method and
     its error calculations over a wide range of input profile
     characteristics and data quality using Monte Carlo simulations.
     We also make comparisons with gas density profiles obtained from
     {\it Chandra} observations, where the PSF correction is small,
     and with profiles obtained using analytic modelling, which
     demonstrate that our procedure is a useful improvement over
     standard techniques. This type of method will be especially
     valuable in the ongoing analysis of unbiased and complete samples
     of X-ray clusters, both local and distant, helping to improve the
     quality of their results.}

\authorrunning{J.H. Croston et al.}
\titlerunning{X-ray density profile deprojection}

   \maketitle
%

\section{Introduction}

The density distribution of X-ray-emitting gas in galaxy clusters is
an important probe of the underlying mass distribution. By assuming
hydrostatic equilibrium and spherical symmetry, the total mass
distribution can be obtained from the gas density profile $\rho_{\rm
gas}(r)$ and temperature profile $T(r)$. The use of galaxy clusters to
test structure formation models and as cosmological probes is reliant
on these techniques. However, observations also suggest that
non-gravitational processes can have important effects on the gas
properties of galaxy clusters (e.g. Arnaud, 2005 for a review), which
raises questions about our understanding of the links between gas and
dark matter properties of clusters. Key diagnostics are provided by
the gas entropy distribution, $S=T/\rho_{\rm gas}^ {2/3}$, which
reflect the specific thermo-dynamical history of the gas. (e.g.
Ponman, Cannon \& Navarro 1999; Ponman, Sanderson \& Finoguenov 2003;
Voit 2005; Pratt, Arnaud \& Pointecouteau 2006). It is therefore
essential to be able to accurately measure the gas distributions of
clusters, not only so as to be able to make accurate inferences about
their dark matter properties, but also to investigate the departures
of the gas properties from the self-similar models expected in the
absence of effects such as cooling and non-gravitational heating and
feedback.

Surface brightness profiles of the X-ray emission from galaxy clusters
have been the primary tool for studying their gas distributions since
cluster data became available from early X-ray missions. Until
recently, it was thought that (with the exception of central cooling
flows) cluster profiles could be adequately fitted with a standard
analytical $\beta$ model profile in surface brightness (e.g. Neumann
\& Arnaud 1999). However, the substantial increase in the data quality
of cluster observations in the era of {\it Chandra} and {\it
XMM-Newton} has revealed that cluster gas distributions are
considerably more complex than expected, e.g. from {\it ROSAT} data. A
variety of increasingly complicated analytical models have been used
to fit cluster surface brightness profiles (e.g. Pratt \& Arnaud 2002;
Vikhlinin et al. 2006) and to calculate the corresponding gas density
profile; however, the need for such a range of models, which are not
generally physically motivated, is unsatisfactory. In addition, using
an analytical model means that full information about real structure
in the cluster profile is lost in converting the profile from surface
brightness to emission measure. 

One alternative is direct deprojection of measured surface brightness
profiles (e.g. Fabian et al. 1981; Kriss et al. 1983; White et al.
1997). If we consider a spherical distribution of cluster gas, with a
density distribution (or emission measure distribution, $S_{emit}(r)$)
consisting of a series of concentric spherical shells of radii
$r_0,r_1,..,r_i,..,r_n$, then using simple geometric considerations
(e.g. McLaughlin 1999) we can calculate a 2-D matrix $[R_{proj}]$
whose elements consist of the contributions of each shell $i$ to the
projected emission measure in a series of annuli on the plane of the
sky $j$ having radii $R_0,R_1,..,R_i,..,R_n$ (which may be the same or
different to the $r_i$), e.g. element $R_{proj,ij}$ is the fraction of
the emission from shell $i$ that is observed in annulus $j$. For the
purposes of this work, we consider a square matrix $R_{proj}$, with
$R_0 = r_0$, $R_1 = r_1$, etc, so that the matrix has dimensions $n
\times n$, where $n$ is the number of bins in the observed surface
brightness profile. The product $[R_{proj}][S_{emit}]$ is then the 2-D
emission measure profile as would be observed by a perfect detector.
We can then calculate a second redistribution matrix, $[R_{PSF}]$ that
takes into account the effect of the instrumental PSF, i.e.
$R_{psf,jk}$ is the fraction of counts from annulus $j$ of the `ideal'
profile that are redistributed by the telescope into annulus $k$ in
the final observed surface brightness profile (we assume that the
energy dependence of the PSF is negligible). The relationship between
an observed $S_X$ profile, $C_{obs}(R)$ and the originating 3D
emission profile, $S_{emit}(r)$ can therefore be expressed as follows:
\begin{equation}
[C_{obs}] = [R_{PSF}][R_{proj}][S_{emit}]
\label{deconv}
\end{equation}
where $C_{obs}(i)$ is the surface brightness detected in an annulus
$i$, $S_{emit}(j)$ is the emission measure produced by a 3-D shell
$j$, $R_{proj}(i,j)$ represents the fraction of emission from shell
$j$ that would be observed by a perfect detector to fall in annulus
$k$, and $R_{PSF}(i,k)$ a second redistribution of counts from an
annulus $k$ of the ideal profile to annulus $i$ of the actual profile
resulting from the effect of the PSF. $R_{proj}$ depends only on the
geometry of the cluster (see e.g. McLaughlin 1999) and can be easily
calculated, and $R_{PSF}$ can also be calculated based on knowledge of
the optical properties of a given instrument. 

However, it is not straightforward to solve Equation~\ref{deconv}
directly, as it is an inverse problem: small amounts of noise in the
data become greatly amplified in the solution for $S_{emit}$. This has
limited the usefulness of such an approach for the analysis of cluster
profiles, particularly for an instrument such as {\it XMM-Newton} with
a reasonably large PSF. The use of ``onion-skin'' deprojection
techniques is common for spectral analysis, and has also been applied
to surface brightness profiles from {\it Chandra} (e.g. David et al.
2001); however, the ``onion-skin'' approach is heavily dependent on
the choice of outermost bin. For Equation~\ref{deconv} to be accurate,
it is necessary to take account of the contribution to each annulus
from shells outside the outermost annulus chosen for the analysis (we
discuss this further in Section~\ref{method}). Finally, it is
necessary to convert the resulting emission measure profiles to gas
density profiles taking into account the variation of the cooling
function $\Lambda(T,Z)$ with radius.

This paper describes the application of a regularisation technique to
the direct deprojection and PSF-deconvolution of X-ray surface
brightness profiles that has allowed us to overcome these limitations.
Our method is motivated by, and adapted from, work by Bouchet (1995)
on the deconvolution of gamma-ray spectra. In the following section,
we describe the regularisation procedure in more detail. We then
present in Section~\ref{simulations} the results of a range of tests
using simulated data that demonstrate the reliability of our technique
and its applicability to a wide range in data quality and cluster
properties. In Section~\ref{data}, we apply our technique to real data
from {\it XMM-Newton} and {\it Chandra} and demonstrate that it
performs well in comparison to other methods. Finally, in
Section~\ref{dlogne}, we investigate the effect of using gas density
profiles obtained from our method on the calculation of the
logarithmic slope and thus total mass profiles for clusters.

\section{Method for regularised deprojection and PSF-deconvolution}
\label{method}
Our method for regularising the deprojection/PSF-deconvolution process
is motivated by the analysis of Bouchet (1995) on the deconvolution of
gamma-ray spectra. The general method is to introduce additional
constraints on the solution of Equation~\ref{deconv} based on prior
information, in this case simply the expectation that the solution
should be smooth, rather than unphysically noisy (see
Sections~\ref{regul} and \ref{lambda}). It is then necessary to introduce a means of
balancing the regularisation constraint with the accuracy to which the
solution reproduces the data (parameterised by $\chi^{2}$). As
described by Bouchet, this is done by using Lagrangian multipliers to
find a solution, then minimising the function
\begin{equation}
L(f,\lambda) = \chi^2(f) + \lambda C(f)
\end{equation}
where $f$ is the solution, $C(f)$ a function that is a minimum when
the solution best satisfies the regularity constraint, and $\lambda$
is the smoothing parameter. By varying $\lambda$, we can therefore
vary the degree to which the solution is dominated by consistency with
the data or with the regularising constraints. The choice of $\lambda$
is therefore critical to obtaining a reliable solution, and is
described further in Section~\ref{lambda}. 

\subsection{The regularisation constraint}
\label{regul}
Again following Bouchet (1995), we adopt the Philips-Towmey
regularisation method (Phillips 1962; Towmey 1963). Generally, this
method consists of minimizing the sum of the squares of the $k^{th}$
order derivates of the solution around each datapoint. For our
purposes, we define the ``smoothness'' constraint as the minimum in
the deviation of the solution from a constant about each data
point, so that:
\begin{equation}
C(f) = \sum_{j=2}^{N-1} (f_{j-1} + f_{j})^{2}
\end{equation}

Since our data covers several orders of magnitude in surface
brightness, it is necessary to reduce the dynamic range of the problem
by rescaling the data by a rough best-fitting model. Bouchet found
that the choice of scaling function only affected the solution for low
signal-to-noise data or when the model fit to the data was poor. We
initially used a $\beta$ model to scale the data, and found that in
cases where the fit was poor (caused by the data being more centrally
peaked than can be represented by a $\beta$ model) the solution tended
to bias towards a flatter profile; however, this affected only the
central few bins. We then adopted the AB model of Pratt \& Arnaud
(2002), which is a modified version of the $\beta$ model that can
roughly fit both centrally peaked and cored models. This model was
tested on a large sample of clusters and found to achieve adequate
fits in all cases with no systematic residuals at the centre. It is
therefore better suited as a scaling function for our regularisation
procedure, and we demonstrate in Section~\ref{simulations} that our
rescaling procedure using the AB model does not lead to any bias in
the output emission measure and gas density profiles. The
regularisation constraint is therefore applied to the rescaled data.

\begin{figure*}
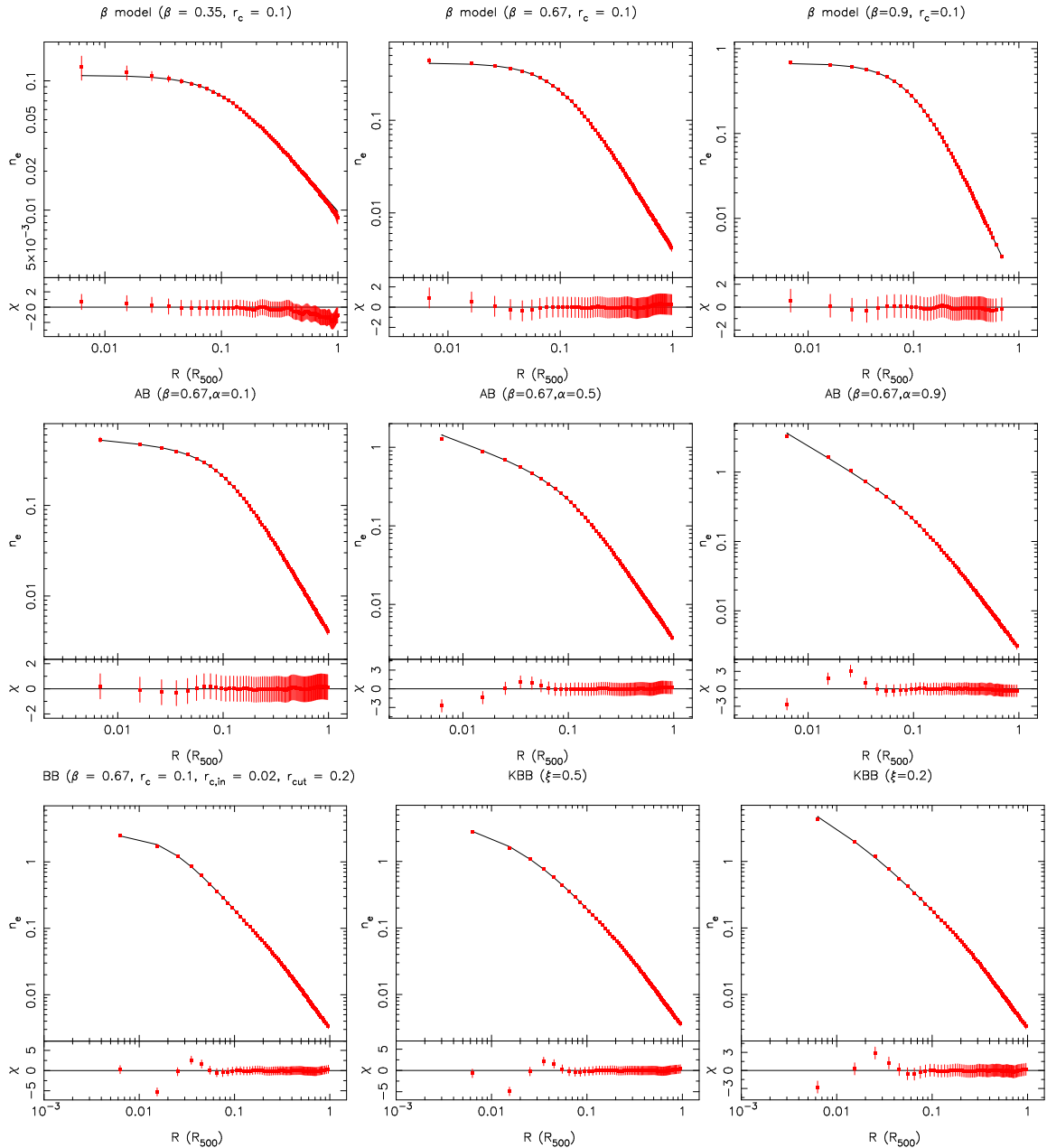

\centerline{
\vbox{
\hbox{
\epsfig{figure=beta_0.35_0.1.ps,width=5.0cm}
\epsfig{figure=beta_0.67_0.1.ps,width=5.0cm}
\epsfig{figure=beta_0.9_0.1.ps,width=5.0cm}}
\hbox{
\epsfig{figure=ab_0.67_0.1_0.1.ps,width=5.0cm}
\epsfig{figure=ab_0.67_0.1_0.5.ps,width=5.0cm}
\epsfig{figure=ab_0.67_0.1_0.9.ps,width=5.0cm}}
\hbox{
\epsfig{figure=bb_0.67_0.1.ps,width=5.1cm}
\epsfig{figure=kbb_xi0.5.ps,width=5.0cm}
\epsfig{figure=kbb_xi0.2.ps,width=5.0cm}}
}}
\caption{Results of Monte Carlo simulations of the deprojection and
   PSF deconvolution of surface brigthness profiles corresponding to
   different shapes of the input model density profiles, as indicated
   in the individual plot labels. The simulated
   surface brightness profiles have a global S/N ratio of 200, for a
   total source/background count ratio of $R=0.3$. In all cases, the
   mean output density profile and errors are shown by red squares,
   with the input density model profile indicated by a black solid
   line.}
\label{profiles}
\end{figure*}

\begin{figure*}
\centerline{
\vbox{
\hbox{
\epsfig{figure=beta0.35_0.1_dlogne.ps,width=5.0cm}
\epsfig{figure=beta0.67_0.1_dlogne.ps,width=5.0cm}
\epsfig{figure=beta0.9_0.1_dlogne.ps,width=5.0cm}}
\hbox{
\epsfig{figure=ab0.67_0.1_0.1_dlogne.ps,width=5.0cm}
\epsfig{figure=ab0.67_0.1_0.5_dlogne.ps,width=5.0cm}
\epsfig{figure=ab0.67_0.1_0.9_dlogne.ps,width=5.0cm}}
\hbox{
\epsfig{figure=bb_0.67_0.1_dlogne.ps,width=5.1cm}
\epsfig{figure=kbb_xi0.5_dlogne.ps,width=5.0cm}
\epsfig{figure=kbb_xi0.2_dlogne.ps,width=5.0cm}}
}}
\caption{Same simulations as for Fig.~\ref{profiles}. The mean output
  profile of $d\log{n_{e}}/d\log{r}$ and errors are shown by red
  squares, with the profile obtained from the input density model
  indicated by a black solid line.}
\label{profilesdlogne}
\end{figure*}

\begin{figure*}
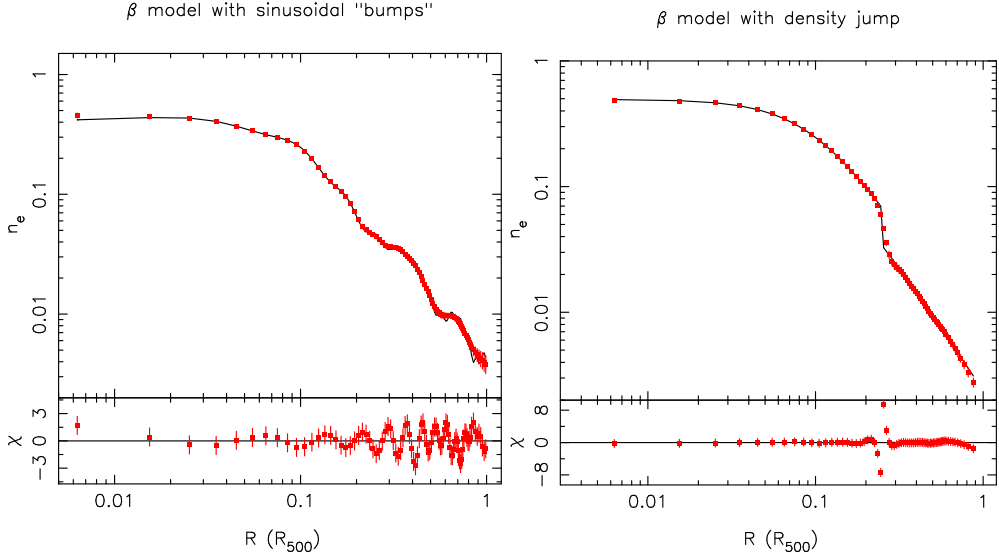

\centerline{
\hbox{
\epsfig{figure=bumpy_sin.ps,width=6.6cm}
\epsfig{figure=coldfront.ps,width=6.5cm}}}
\caption{Monte Carlo simulations of the deprojection and PSF
  deconvolution of ``bumpy'' surface brightness profiles. The lefthand
  density profile is a beta model with sinusoidal variations
  superimposed, and the righthand profile is a beta model with a
  density jump, such as might be produced by a cold front. Red symbols
  are the output profile; solid black line is the input model.}
\label{bumpy}
\end{figure*}

\begin{figure*}
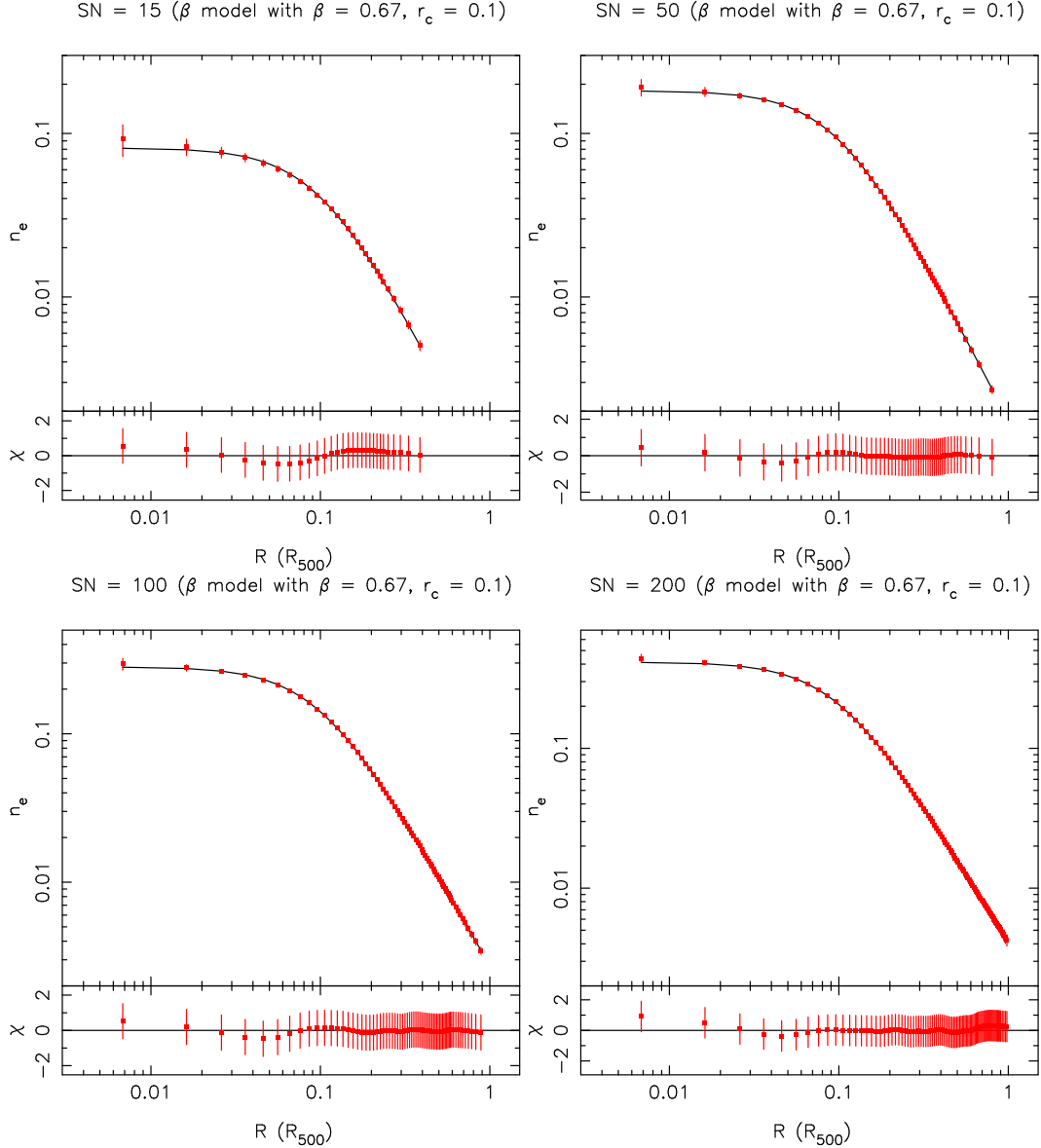

\centerline{
\vbox{
\hbox{
\epsfig{figure=sn15.ps,width=7.0cm}
\epsfig{figure=sn50.ps,width=7.0cm}}
\hbox{
\epsfig{figure=sn100.ps,width=7.0cm}
\epsfig{figure=sn200.ps,width=7.0cm}}
}}
\caption{Results of Monte Carlo simulations of the deprojection and
   PSF deconvolution of surface brigthness profiles with different
   global S/N ratios and source/background count ratios $R$ ($3\sigma$
   binning in cases), as indicated in the individual plot labels. In
   all cases, the mean output density profile and errors are shown by
   red squares, with the input density model profile indicated by a
   black solid line.}
\label{globalsn}
\end{figure*}

\begin{figure*}
\centerline{
\vbox{
\hbox{
\epsfig{figure=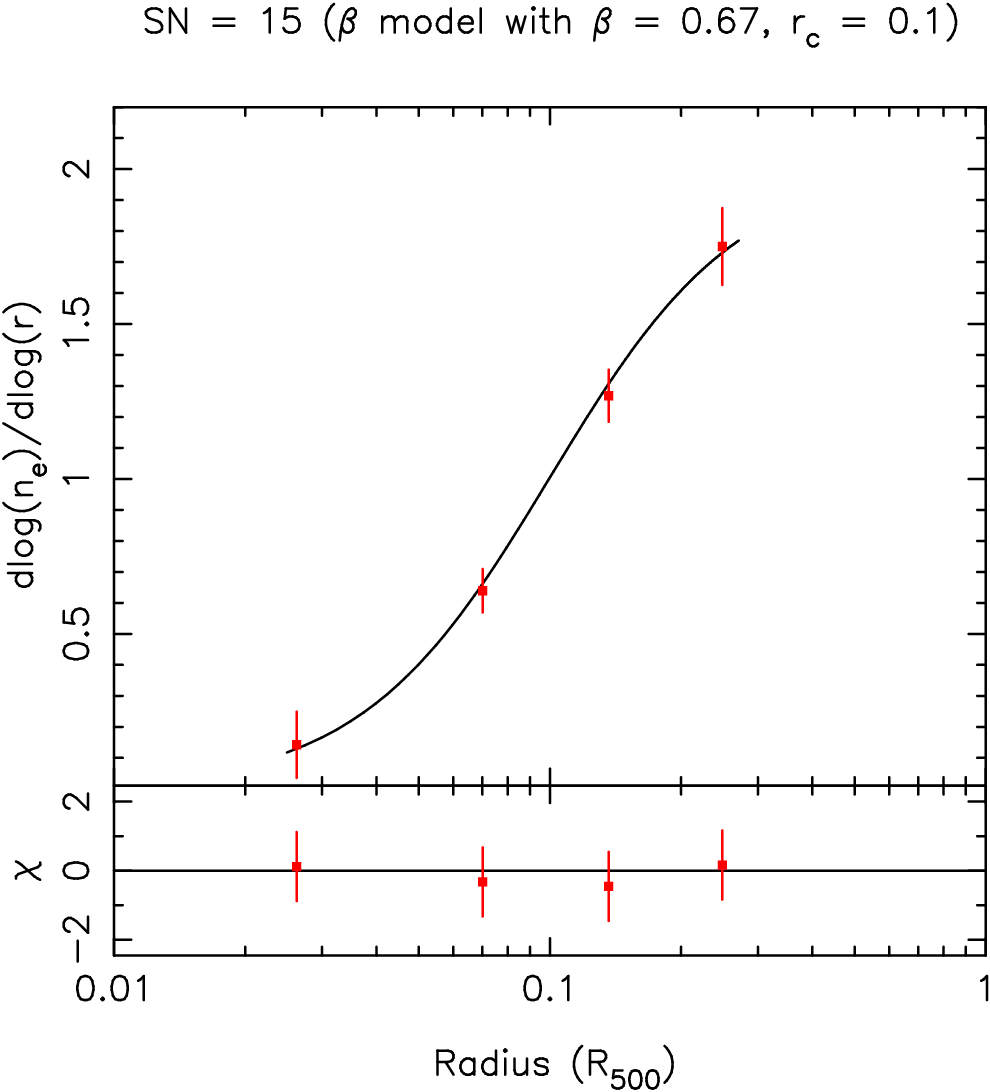,width=7.0cm}
\epsfig{figure=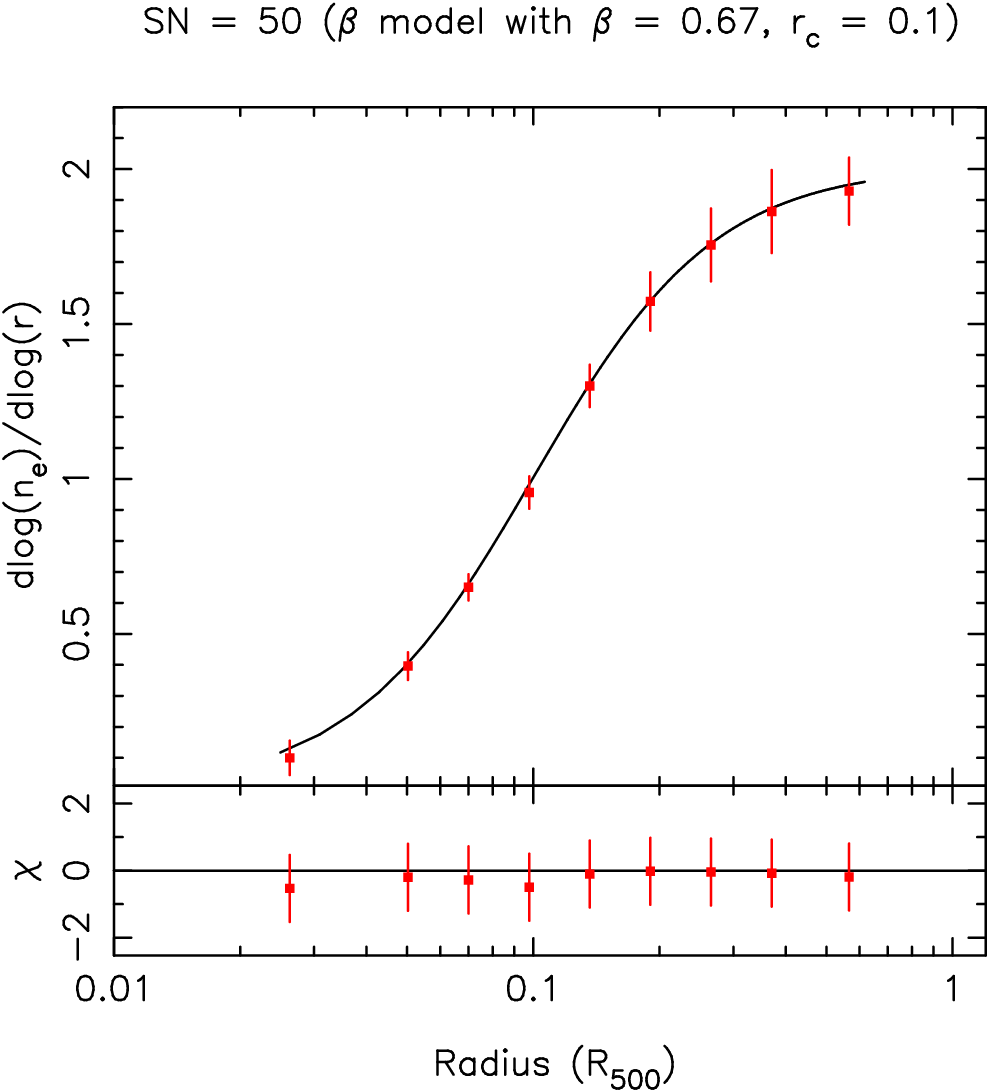,width=7.0cm}}
\hbox{
\epsfig{figure=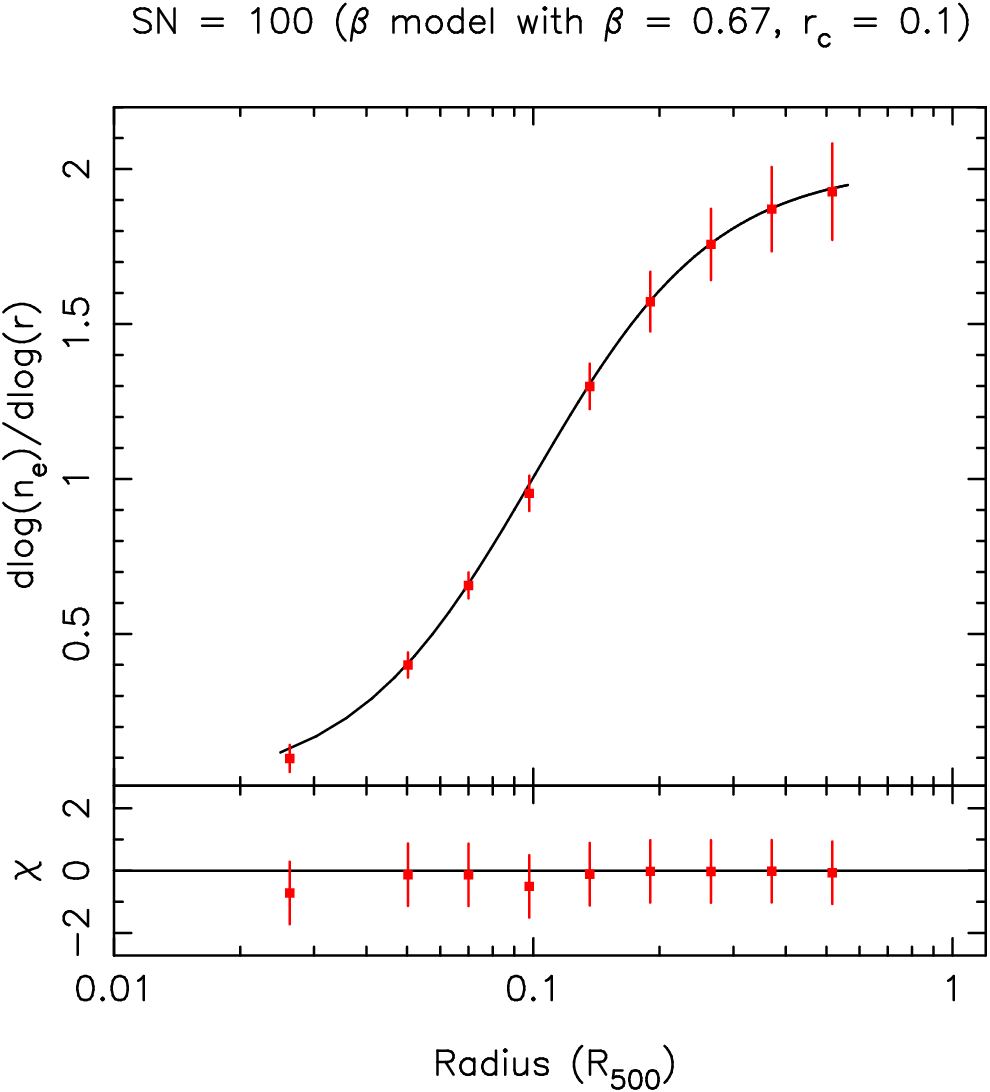,width=7.0cm}
\epsfig{figure=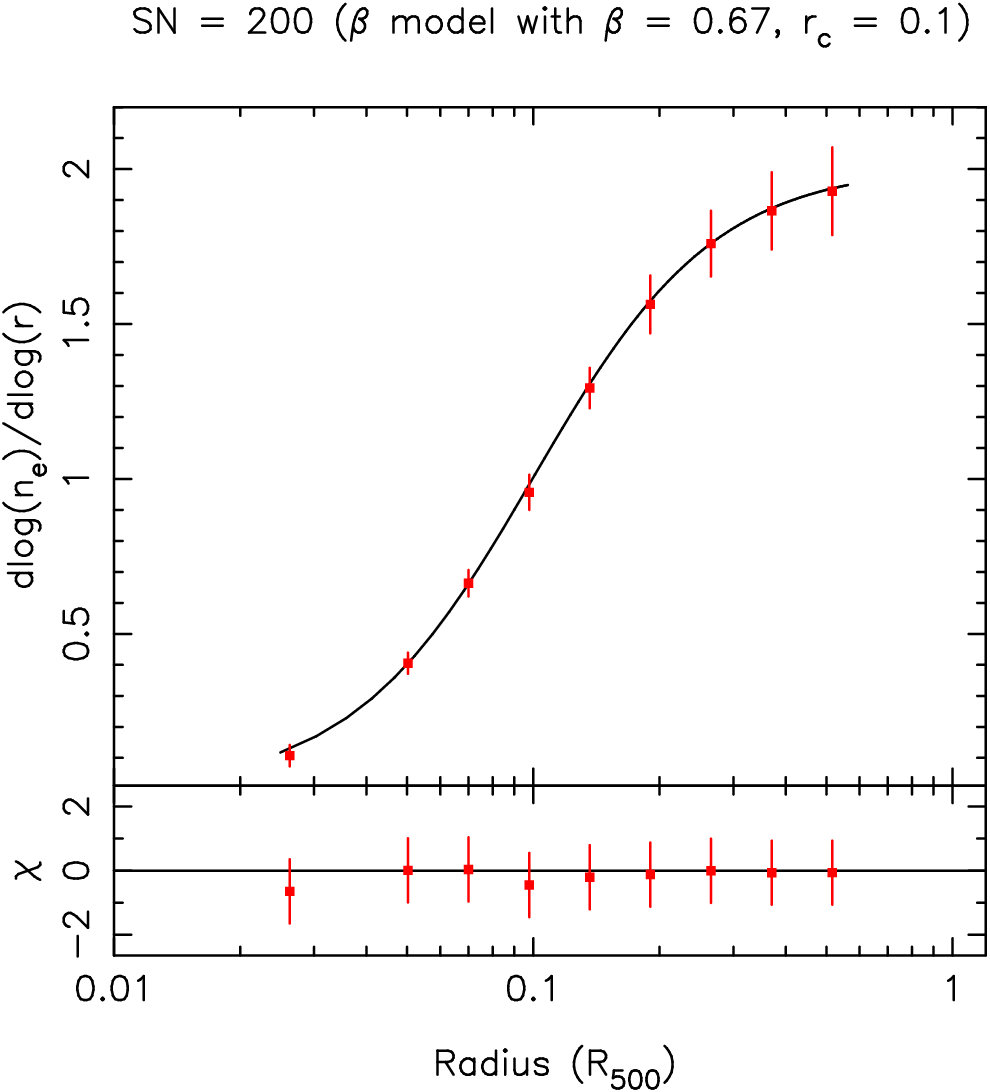,width=7.0cm}}
}}
\caption{Same simulations as Fig.~\ref{globalsn}. The mean output
  profile of $d\log{n_{e}}/d\log{r}$ and errors are shown by red
  squares, with the profile obtained from the input density model
  indicated by a black solid line.}
\label{globalsndlogne}
\end{figure*}

\begin{figure*}
\centerline{
\vbox{
\hbox{
\epsfig{figure=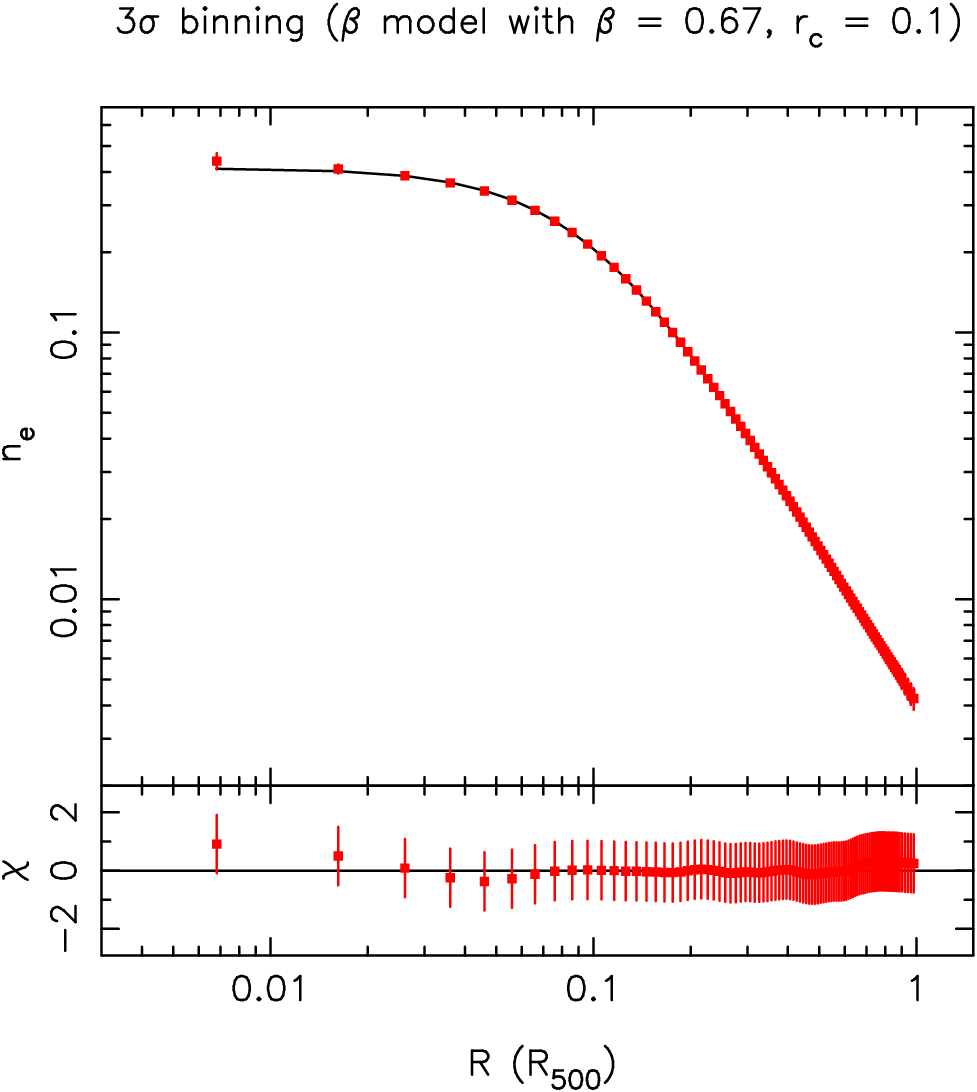,width=7.0cm}
\epsfig{figure=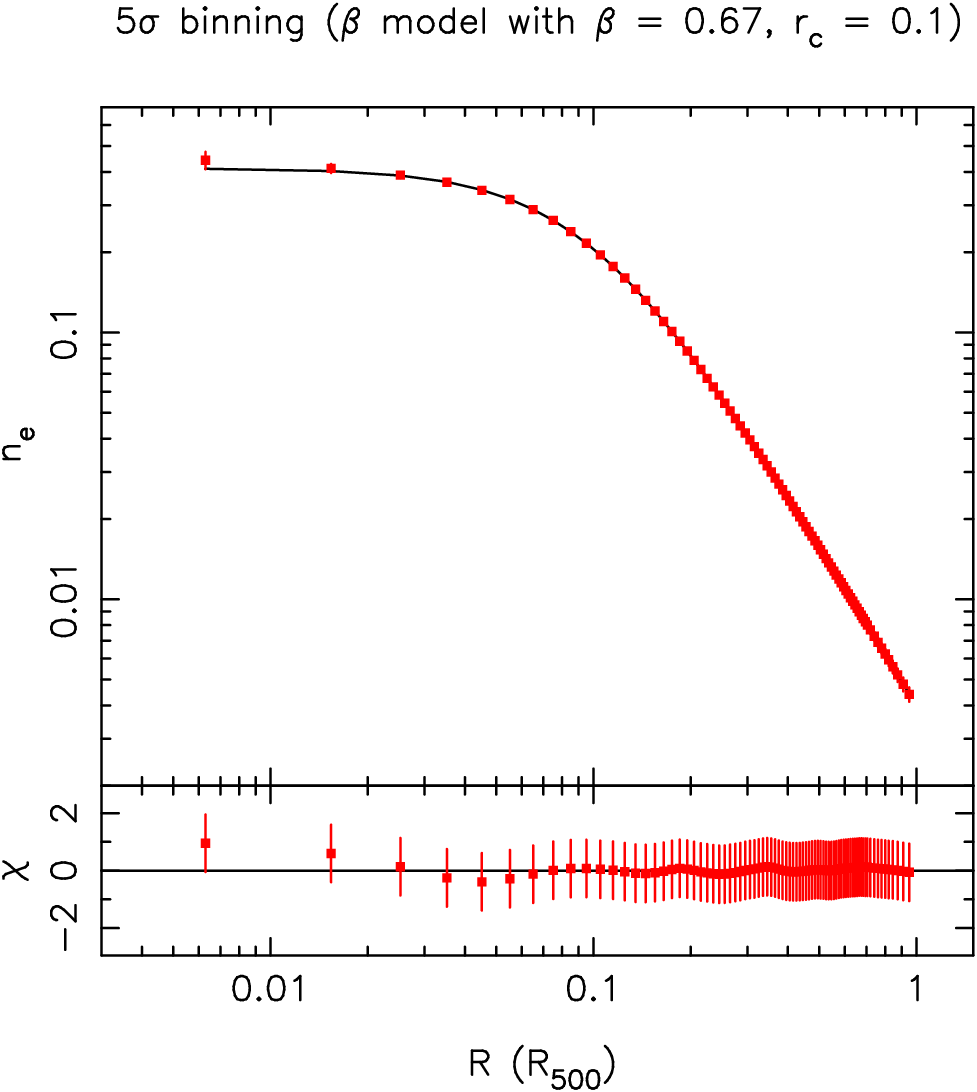,width=7.0cm}}
\hbox{
\epsfig{figure=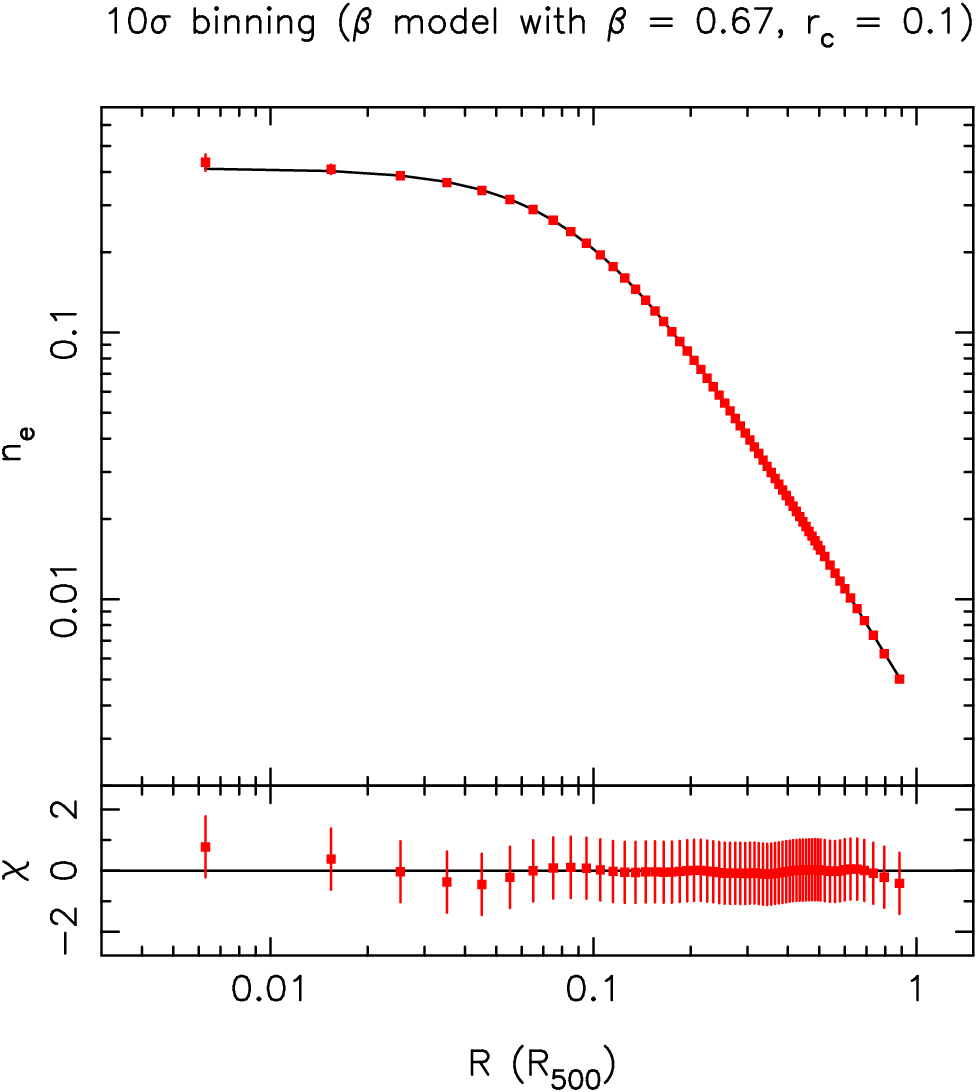,width=7.0cm}
\epsfig{figure=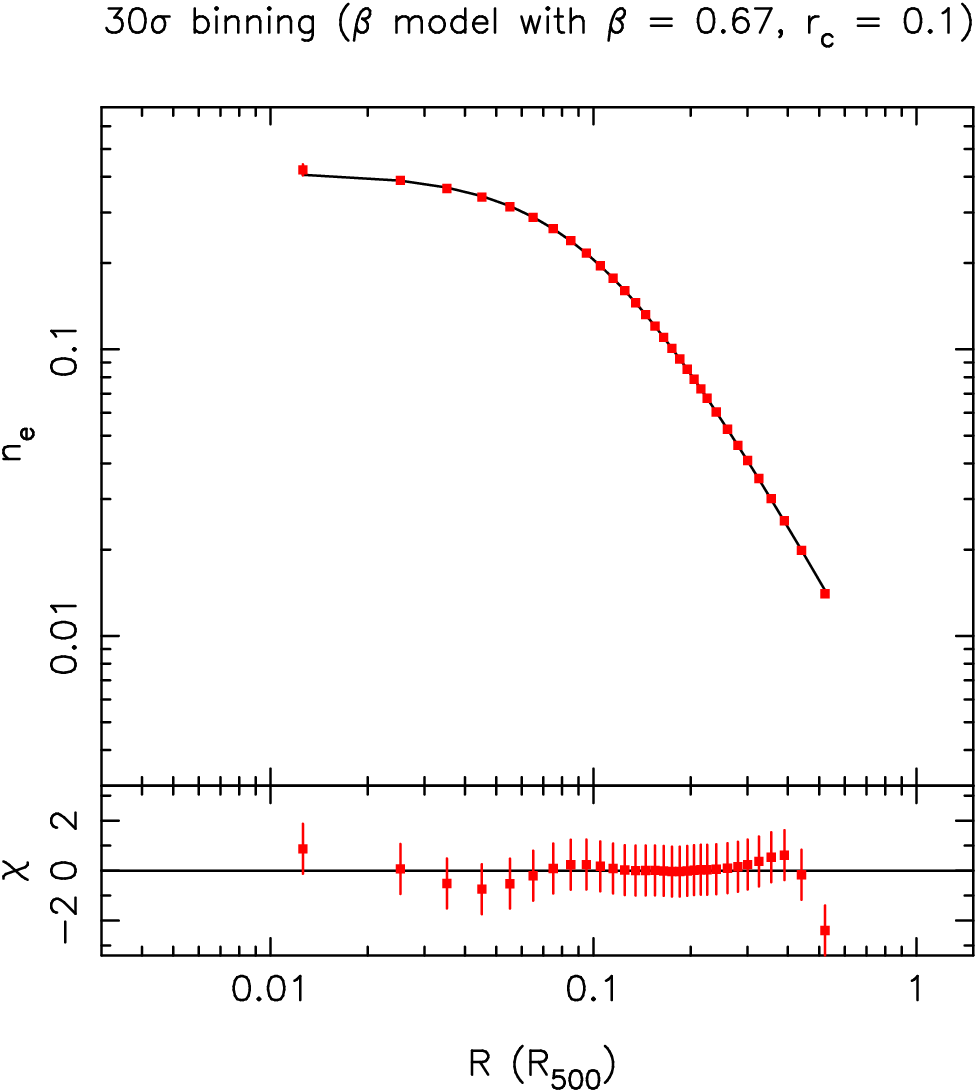,width=7.0cm}}
}}
\caption{Results of Monte Carlo simulations of the deprojection and
   PSF deconvolution of surface brigthness profiles, using different
   binning of the data (signal-to-noise per bin), as indicated in the
   individual plot labels. The simulated surface brigthness profiles
   have a global S/N ratio of 200, for a total source/background count
   ratio of $R=0.3$. In all cases, the mean output density profile and
   errors are shown by red squares, with the input profile indicated
   by a black solid line.}
\label{snperbin}
\end{figure*}

\begin{figure*}
\centerline{
\vbox{
\hbox{
\epsfig{figure=snpb3_dlogne.ps,width=7.0cm}
\epsfig{figure=snpb5_dlogne.ps,width=7.0cm}}
\hbox{
\epsfig{figure=snpb10_dlogne.ps,width=7.0cm}
\epsfig{figure=snpb30_dlogne.ps,width=7.0cm}}
}}
\caption{Same simulations as Fig.~\ref{snperbin}. The mean output
  profile of $d\log{n_{e}}/d\log{r}$ and errors are shown by red
  squares, with the profile obtained from the input density model
  indicated by a black solid line.}
\label{snpbdlogne}
\end{figure*}

\begin{figure*}
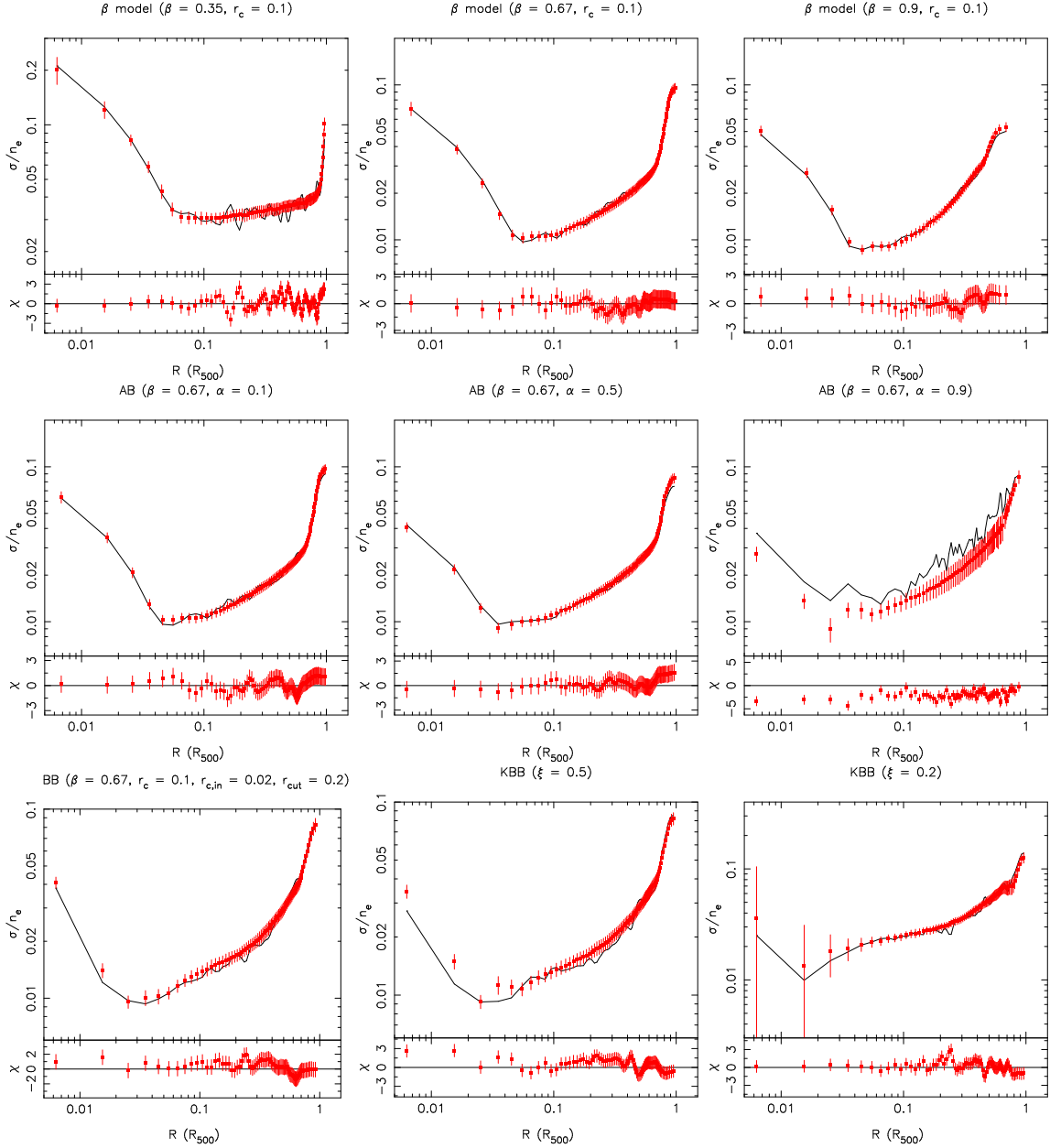

\centerline{
\vbox{
\hbox{
\epsfig{figure=beta_0.35_0.1_err.ps,width=5.0cm}
\epsfig{figure=beta_0.67_0.1_err.ps,width=5.0cm}
\epsfig{figure=beta_0.9_0.1_err.ps,width=5.0cm}}
\hbox{
\epsfig{figure=ab_0.67_0.1_0.1_err.ps,width=5.0cm}
\epsfig{figure=ab_0.67_0.1_0.5_err.ps,width=5.0cm}
\epsfig{figure=ab_0.67_0.1_0.9_err.ps,width=5.0cm}}
\hbox{
\epsfig{figure=bb_0.67_0.1_err.ps,width=5.0cm}
\epsfig{figure=kbb_xi0.5_err.ps,width=5.0cm}
\epsfig{figure=kbb_xi0.2_err.ps,width=5.0cm}}
}}
\caption{Error estimation from Monte Carlo simulations of the
  deprojection and PSF deconvolution of input model density profiles
  of different profile shapes. In all cases, the relative mean ouput
  errors are shown by red squares, with the ``true'' errors indicated
  by a black solid line (the ``true'' error profile can be made
  smoother by increasing the number of MC iterations).}
\label{profileserr}
\end{figure*}

\begin{figure*}
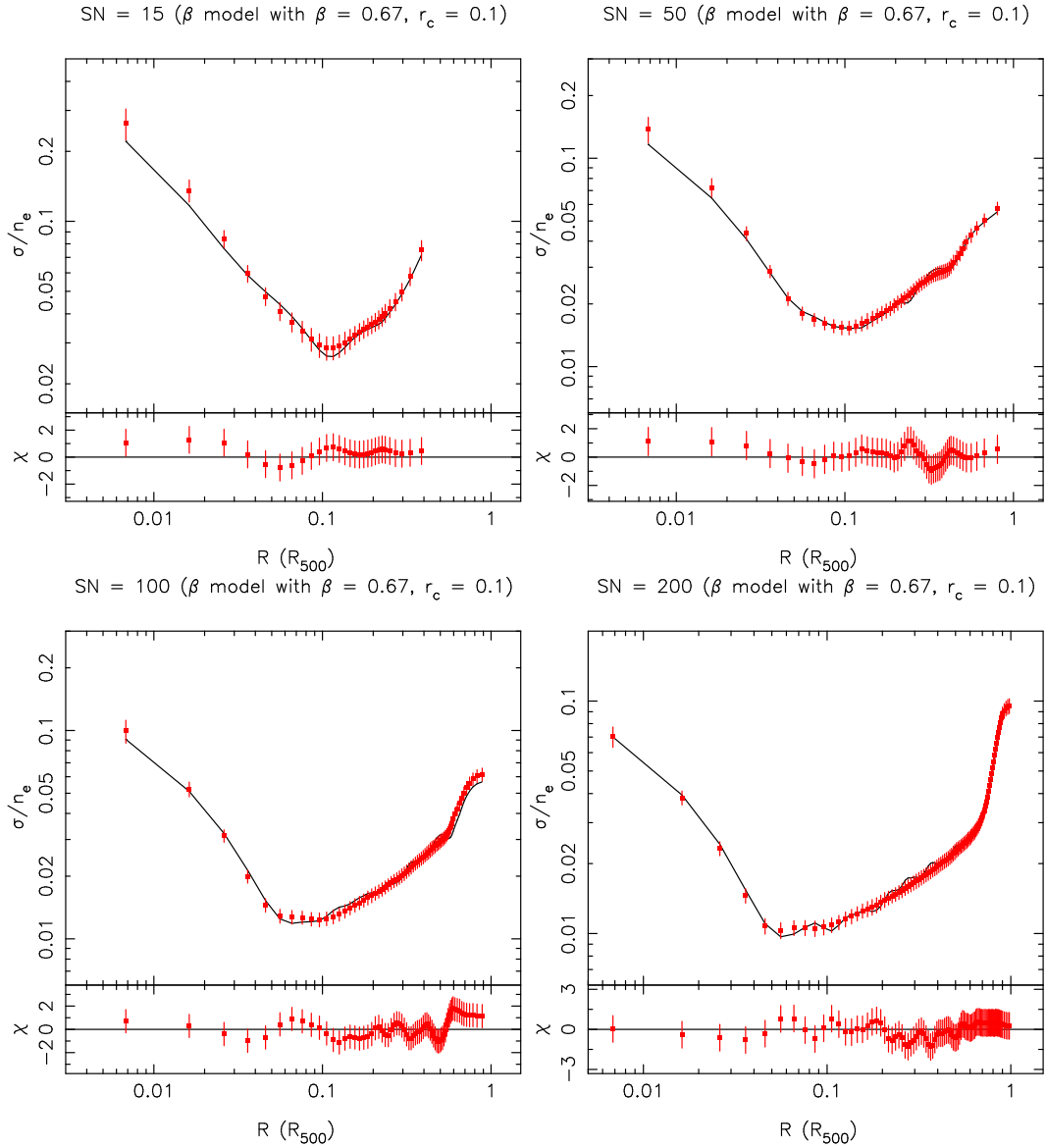

\centerline{
\vbox{
\hbox{
\epsfig{figure=sn15_err.ps,width=7.0cm}
\epsfig{figure=sn50_err.ps,width=7.0cm}}
\hbox{
\epsfig{figure=sn100_err.ps,width=7.0cm}
\epsfig{figure=sn200_err.ps,width=7.0cm}}
}}
\caption{Error estimation from Monte Carlo simulations of the
  deprojection and PSF deconvolution of input model density profiles
  of different global S/N ratios ($3\sigma$ binning in cases). In all
  cases, the relative mean output errors are shown by red squares,
  with the ``true'' errors indicated by a black solid line.}
\label{globalsnerr}
\end{figure*}

\begin{figure*}
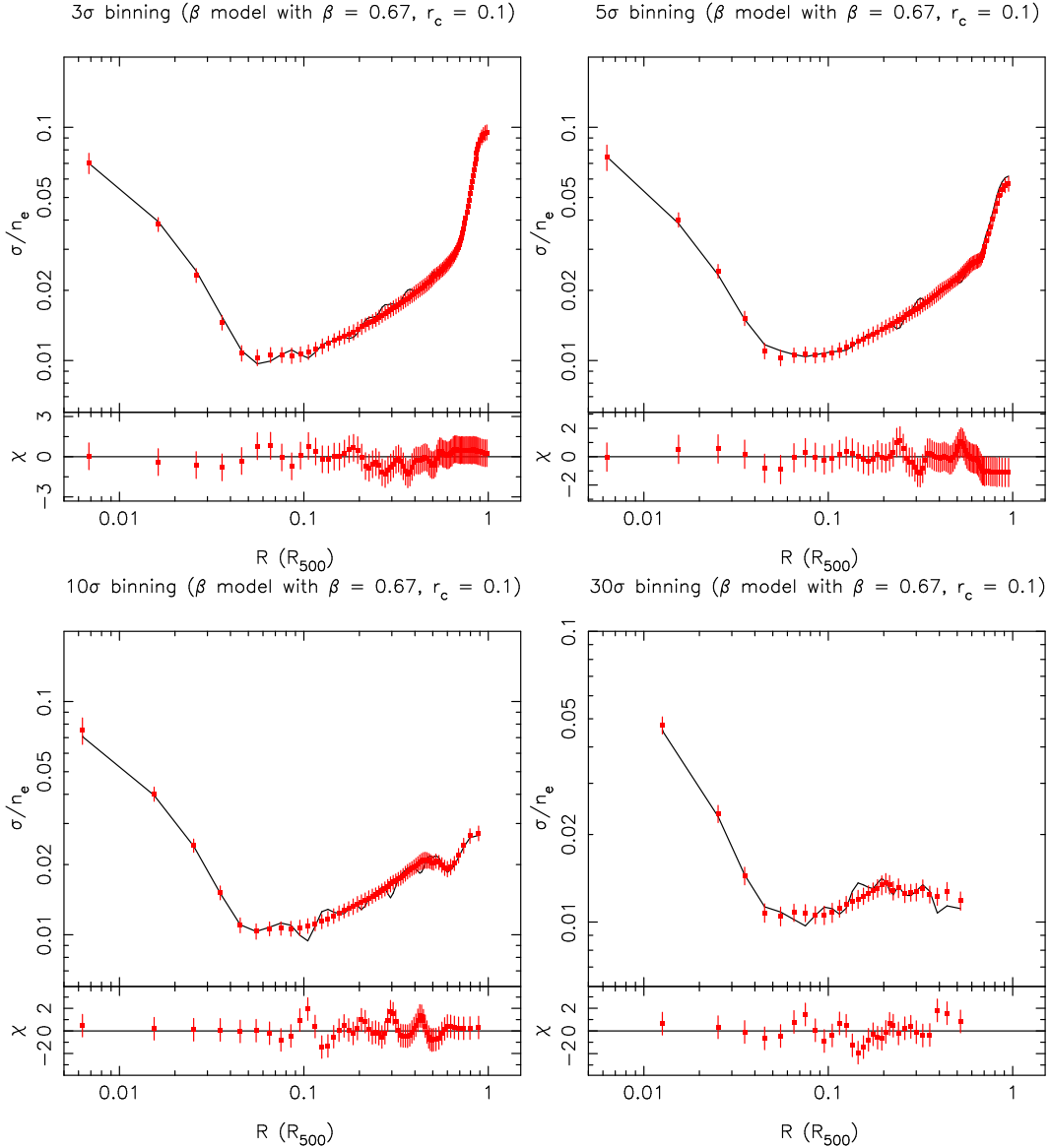

\centerline{
\vbox{
\hbox{
\epsfig{figure=snpb3_err.ps,width=7.0cm}
\epsfig{figure=snpb5_err.ps,width=7.0cm}}
\hbox{
\epsfig{figure=snpb10_err.ps,width=7.0cm}
\epsfig{figure=snpb30_err.ps,width=7.0cm}}
}}
\caption{Error estimation from Monte Carlo simulations of the deprojection and
  PSF deconvolution of input model density profiles of different
  signal-to-noise per bin (for global S/N = 200). In all cases, the
  relative mean output errors are shown by red squares,
  with the ``true'' errors indicated by a black solid line.}
\label{snperbinerr}
\end{figure*}

\subsection{Choice of the smoothing parameter}
\label{lambda}
As mentioned above, the smoothing parameter, $\lambda$, is used to
achieve a correct balance between fidelity to the input data and
consistency with the applied regularisation constraints. It is
therefore critical to adopt an objective means for choosing
$\lambda$. Again following Bouchet (1995), we implemented a
cross-validation technique (e.g. Wahba 1978) that consists of
predicting each datapoint by finding a solution using all the data
excluding that point. For a value of $\lambda$ that is too low
(i.e. insufficient smoothing), cross-validation will poorly predict
the data because rapidly varying solutions will not accurately
represent the excluded datapoints. In constrast, for a value of
$\lambda$ that is too high (i.e. oversmoothing), again the data will
be poorly predicted by cross-validation, because real, larger-scale
variations in the data will have been lost. Cross-validation therefore
offers a systematic, objective means of choosing a best value for the
smoothing parameter. We implement this method in the same way as
Bouchet (1995).

\subsection{Error calculation}
\label{errs}

To calculate the errors on the output emission measure profile, we
initially used the standard Monte Carlo technique of adding Gaussian
noise to the observed profile to generate 100 profiles, applying the
code and using the dispersion in the output emission measure for each
bin as the error. However, we found that this method tends to
overestimate the true errors, because it adds noise to an already
noisy profile. Instead, we adopted a more complicated technique, which
was found to give errors that are unbiased, as follows:
\begin{enumerate}
\item The output emission measure profile is fitted with an AB model.
\item The corresponding $S_X$ profile is calculated.
\item Gaussian noise is added to the model $S_X$ profile to generate
  100 simulated profiles.
\item The simulated $S_X$ profiles are deprojected.
\item The dispersion (standard deviation) in the distribution of the
  output emission measure for each bin is adopted as the error for
  that bin.
\end{enumerate}
The use of a model distribution avoids the problem of adding in an
additional scatter from the noise in the original observed profile.
This error estimation technique was tested as part of the simulations
described in Section~\ref{simulations}.

\subsection{Calculation of the density slope}
\label{slopecalc}
For the purposes of obtaining total mass profiles, it is particularly
important that any method for obtaining density profiles accurately
calculates the logarithmic density slope ($d\log{n_e}/d\log{r}$). In
principle it is possible to calculate the slope for each radial bin of
the input surface brightness profile; however, it is necessary to use
the Monte Carlo method to obtain the error on the slope at each point
as the errors on density are likely to be correlated, and we found
that although the mean $d\log{n_e}/d\log{r}$ profile over 100
simulations is well recovered, small variations in the profile shape
lead to very large error bars on the slope. In practise, the
calculation of total mass will always be limited by the
signal-to-noise achievable in the temperature profile. To obtain good
constraints on $d\log{n_e}/d\log{r}$, we therefore decided to carry
out the slope calculation in larger radial bins, corresponding to
those used for the cluster temperature profile (for the tests carried
out here, we used binning typical of an {\it XMM-Newton} temperature
profile for a cluster observation of the appropriate signal-to-noise
ratio, which corresponds to bins roughly ten times larger than the
surface brightness bins). The slope was then calculated for each bin
using a linear least-squares fit to all of the density data points
falling in that bin. We adopted the dispersion (standard deviation) of
the calculated slope at each radius from the MC simulations as for the
errors on $n_{e}$. This method for obtaining $d\log{n_e}/d\log{r}$ was
also tested as part of our simulations and using X-ray data, as
described in the following sections.

\subsection{Correcting for X-ray emission beyond the profile region}

When transforming an observed surface brightness profile to an
emission measure profile, there will be a contribution to each
observed surface brightness bin from emission at larger radii than the
outermost profile bin. When fitting analytical models, this
contribution is taken into account, since the surface brightness
models are obtained by integrating along the line of sight; however,
with our deprojection method it is necessary to correct for this
contribution before carrying out the deprojection. We use the method
described by McLaughlin (1999) (his equation A4, with the slope
$\alpha$ measured from the data), which uses the assumption that
$S_{X} \propto r^{-\alpha}$ at large radius to subtract the
contribution from this emission to each profile bin.

\section{Monte Carlo simulations}
\label{simulations}
In order to test how accurately the deprojection code recovers the
correct density profile, and how well the error calculation represents
the true uncertainty, we carried out a series of tests using Monte
Carlo simulations. Here we neglect the variation of temperature and
abundance with radius, and assume a one-to-one relation between the
emission measure and the gas density. A model density profile was used
to calculate corresponding surface brightness profiles including
Gaussian noise (100 surface brightness profiles were generated for
each model). We assumed a flat background, which was used to define
the global signal-to-noise of the profile, and to calculate realistic
errors as a function of radius. We chose representative values (see
below) for the global signal-to-noise and the ratio ($R$) between
total source and background counts for different signal-to-noise
profiles based on a sample of nearby clusters (Pointecouteau et al.
2005), and a sample of distant clusters (Arnaud et al., in prep.),
both observed with {\it XMM-Newton}. If $N_i$ and $N_{bg}$ are the net
source and assumed background counts in each bin, the error $\sigma_i$
on the estimated source counts after background subtraction is the
quadractic sum of the error on the observed counts before background
subtraction and the error on the estimated background counts in the
extraction region. We consider the conservative case where this error
is simply $\sqrt{N_{bg}}$. The errors, $\sigma_i$, were thus calculated
according to\footnote{In cases where the background level
  can be estimated from blank-sky fields with long exposures or large
  extraction regions, the background may be better determined than the
  source, in which case this expression will overestimate the errors.}:
\begin{equation}
\sigma_{i} = \sqrt{N_i + 2.0N_{bg}}
\end{equation}
on the assumption of a flat background of total counts $N_{btot} = R
N_{tot}$ within $R_{500}$.

Each surface brightness profile was then convolved with a PSF matrix:
we used a typical {\it XMM-Newton} redistribution matrix based on the
analytical model of Ghizzardi et al. (2001) and assuming that the
surface brightness profiles contains contributions from all 3 {\it
XMM-Newton} cameras (with weights of 0.25,0.25,0.5 for the MOS1, MOS2
and pn cameras, respectively). The {\it XMM-Newton} PSF was used so as
to test the code in situations where the PSF correction is large
compared to the bin size. To convert between the PSF in arcmin and
units of $R_{500}$, we assumed that $R_{500}$ = 12 arcmin, so that the
PSF FWHM of $\sim 6$ arcsec corresponds to $\sim 1 - 2$ surface
brightness bins. Each simulated profile was then run through the
deprojection code, and the mean and standard deviation of the output
density for each radial bin was tabulated. The logarithmic density
slope ($d\log{n_e}/d\log{r}$) was also compared with the slope of the
input density profile.

Firstly we tested the effect of profile shape on code performance,
using a variety of analytical models. We tested the standard $\beta$ model:
\begin{equation}
n_e(r) = n_0\left(1 +
\frac{r^{2}}{r_{c}^{2}}\right)^{-\frac{3\beta}{2}}
\end{equation}
as well as several other models that can provide better fits to real
cluster data, in particular fitting centrally peaked profiles. These
included a double $\beta$ model:
\begin{eqnarray}
r < r_{cut} && n_e(r) = n_0\left(1 +
\frac{r^{2}}{r_{c,in}^{2}}\right)^{-\frac{3\beta_{in}}{2}} \\
r > r_{cut} && n_e(r) = N\left(1 + \frac{r^{2}}{r_{c}^{2}}\right)^{-\frac{3\beta}{2}}\\
\end{eqnarray}
where $N$ and $\beta_{in}$ are calculated from the other model
parameters so as both $n_e$ and the slope are continuous (formulae in
Pratt \& Arnaud 2002), a modified double $\beta$ model (the KBB model of Pratt \& Arnaud 2002) to fit more centrally peaked profiles:
\begin{eqnarray}
r < r_{cut} && n_e(r) = n_0\left(1 + \frac{r^{2}}{r_{c,in}^{2\xi}}\right)^{-\frac{3\beta_{in}}{2\xi}}\\
r > r_{cut} && n_e(r) = N\left(1 + \frac{r^{2}}{r_{c}^{2}}\right)^{-\frac{3\beta}{2}} \\
\end{eqnarray}
where $\xi$ determines the degree to which the profile is peaked, and a modified single $\beta$ model (the AB model of Pratt \& Arnaud 2002) that has a similar form to the NFW profile allowing it to fit central cusps:
\begin{equation}
n_{e} (r) = A \left(\frac{r}{r_{c}}\right)^{-\alpha} \left( 1 + \frac{r^{2}}{r_{c}^{2}}\right)^{\frac{-3\beta}{2} + \frac{\alpha}{2}}
\end{equation}
We ran the simulation for each of these profile shapes for several
sets of parameters to test the influence of profile shape on code
performance. The results are shown in Figure~\ref{profiles}. For these
tests, the input surface brightness profiles had a global
signal-to-noise ratio of 200, representative of a typical observation
of a nearby cluster, a total source/background count ratio $R = 0.3$,
appropriate for this signal-to-noise, and were binned to obtain a
signal-to-noise per bin of at least $3\sigma$.

Fig.~\ref{profiles} shows that our deprojection method is not
significantly affected by the profile shape: in all cases the input
density profile is recovered to a high degree of accuracy. A more
quantitative comparison is given in Table~\ref{results}: the reduced
$\chi2$ values for the mean output density profile compared to the
input profile show that there is no significant bias. It appears that
the scaling function does have a small effect on the code performance:
the BB and KBB models, for which the goodness of fit of the scaling AB
model were poorest, have significant residuals in the central few
bins. However, the overall agreement between the mean output and input
profiles (see Table~\ref{results}) is still excellent. In
Figure~\ref{profilesdlogne}, we show the $d\log{n_e}/d\log{r}$
profiles for each of the models, calculated as described in
Section~\ref{slopecalc}, compared with the slope of the input density
profile, calculated directly from the model. As indicated in
Table~\ref{results}, our method accurately and without bias recovers
$d\log{n_e}/d\log{r}$ for all tested input density models (although
the slope is less well-recovered in the innermost bin for the steepest
density profiles).

As one of the main strengths of this method is that it retains full
information from the surface brightness profile, we also wanted to
test its ability to recover information about deviations from a smooth
functional form in the input density profile. We therefore simulated
density profiles containing sinusoidal modulations superimposed on a
$\beta$-model form in order to represent more complex profile
behaviour. Figure~\ref{bumpy} shows the simulation results for one
such profile, which is representative of the ``bumpy'' profiles that
were investigated. Again, the deprojection code recovers extremely
well the shape of the ``bumps'' in the input profile, although they
are slightly smoothed, as is evident in the residual plot. We also
simulated a profile with a density discontinuity (by reducing the
$\beta$ model normalisation by a factor of 2 at a radius of $2.5
r_{c}$) such as might be produced by a cold front. Again, the profile
is well reproduced, although the output density discontinuity is
slightly smoothed compared to the input profile. The slight smoothing
effect will be more severe for very narrow, spiked features (as was
noted by Bouchet (1995) for his original application of this technique
to spectral deconvolution); however, in the case of X-ray surface
brightness profiles, such features would not be expected as they are
unphysical, so that this is not an important limitation of the code.
Although we have demonstrated that radial imhomogeneities in gas
distribution are well recovered by our method, it is important to be
aware of the inherent limitations of a one-dimensional approach to
measuring gas density. Like all methods based on azimuthally symmetric
radial surface brightness profiles, our deprojection method does not
take into account azimuthal variations or variations along the
line-of-sight. Our method could be generalized to ellipsoid shells;
however, this would require assumptions about the cluster structure
along the line-of-sight. A detailed discussion of the accuracy of 1-D
deprojection methods is beyond the scope of this paper.

We next used a $\beta$ model profile (for simplicity) to test the
effects of global signal-to-noise ratio and choice of binning. We
tested $\beta$ model profiles with $\beta = 0.67$ (typical of cluster
profiles) and $r_{c} = 0.1R_{500}$ for global signal-to-noise ratios
of 200, 100, 50 and 15, which fully sample the range in data quality
seen in observations of nearby and distant clusters. The corresponding
total source/background count ratios were 0.3, 1, 2 and 5,
respectively, as obtained using the relationship between
signal-to-noise and $R$ in the observations. Figure~\ref{globalsn}
shows the deprojection results for the four choices of global S/N,
with goodness of fit information in Table~\ref{snresults}.
Unsurprisingly, the code performs best for the highest quality data;
however, in all cases the input density profile is well recovered, and
even at the lowest S/N ratio of 15, the deprojection method performs
well ($\chi^{2}$ of 0.1 for 24 d.o.f. for the mean output profile).
This is further illustrated by the plots of $d\log{n_e}/d\log{r}$
shown in Figure~\ref{globalsndlogne}.

We then tested the effect of binning of the input surface brightness
profile on the performance of our method. Using the same $\beta$ model
parameters as for the global S/N tests, and a global S/N of 200 again,
we tested binning with ratios of 3, 5, 10 and 30$\sigma$ per bin. The
deprojection results for these tests are shown in
Figure~\ref{snperbin}, with goodness of fit information in
Table~\ref{snresults}, and plots of $d\log{n_e}/d\log{r}$ in
Figure~\ref{snpbdlogne}. These figures show that the choice of binning
does not have an important effect on the recovery of the input density
profile.

In addition to testing how well the input density profiles were
recovered in our simulations, we also tested the accuracy of our error
calculation method (described in Section~\ref{errs}). The ``true''
error on the density calculated at a given radius should be given by
the standard deviation of the distribution of density values
($\sigma_{n_{e}}$) obtained over the 100 Monte Carlo runs for a given
model (indeed, this is precisely the method we are using in the error
calculation). We therefore compared the mean errors calculated by our
code at each density ($<\sigma_{calc}>$) with the ``true'' error from
the simulated density distribution ($\sigma_{n_{e}}$) to confirm the
accuracy of our method. Figure~\ref{profileserr} shows a comparison of
$<\sigma_{calc}>(r)$ and $\sigma_{n_e}(r)$ for a range of models, with
Figs.~\ref{globalsnerr} and ~\ref{snperbinerr} illustrating the effect
of global S/N and choice of binning on the accuracy of error
determination. In all cases the errors calculated by the code are
reasonably accurate (within 2$\sigma$ of the true error) and generally
not biased in any systematic way (although for the steepest AB model
there is a slight systematic underestimation, the origin of which is
unclear). The accuracy of the error calculations is also shown in
Table~\ref{results}.

\begin{table*}
\begin{minipage}[t]{\columnwidth}
\caption{Code performance for different profile shapes}            
\label{results}
\centering
\renewcommand{\footnoterule}{}
\begin{tabular}{l l l l l l l l l l l l}       
\hline\hline                
Model & $\beta$ & $r_{c}$ & $r_{c,in}$ & $r_{cut}$ & $\alpha$ &
$\xi$ & $\chi^{2}_{dens}/N_{bins}$\footnote{Using mean profile from
  100 simulation runs} &$\chi^{2}_{denserrs}/N_{bins}$\footnote{Comparison of mean
  errors with `true' errors -- see text} & $\chi^{2}_{slope}/N_{bins}$\footnote{Using mean logarithmic density slope from 100 simulation runs}&$\chi^{2}_{slopeerrs}/N_{bins}$\\   
\hline                      
$\beta$ & 0.67 & 0.1 & -- & -- & -- & -- & 1.54/55 & 45.2/55& 2.7/54 & 39.5/54\\
$\beta$ & 0.35 & 0.1 & -- & -- & -- & -- & 23.0/64 & 76.2/63 & 4.5/63 & 52.1/60\\
$\beta$ & 0.9 & 0.1 & -- & -- & -- & -- & 5.5/47 & 39.2/47 & 4.2/46 & 30.7/46\\
AB & 0.67 & 0.1 & -- & -- & 0.1 & -- & 1.02/54 & 84.2/54 & 1.1/54 & 94.0/54\\
AB & 0.67 & 0.1 & -- & -- & 0.5 & -- & 2.65/52 & 42.5/52 & 9.2/52 & 26.5/52\\
AB & 0.67 & 0.1 & -- & -- & 0.9 & -- & 8.46/49 & 229/48 & 5.9/49 & 228/46\\
BB & 0.67 & 0.1 & 0.02 & 0.2 & -- & -- & 19.1/49 & 31.1/49 & 36.1/49 & 39.3/49\\
KBB & 0.67 & 0.1 & 0.02 & 0.2 & -- & 0.5 & 11.8/50 & 71.9/50 & 19.4/50 & 67.0/50\\
KBB & 0.67 & 0.1 & 0.02 & 0.2 & -- & 0.2 & 2.92/49 & 56.8/49 & 5.1/49 & 39.5/49\\
bumpy & - & - & - & - & - & - & 111.3/91 & - & 121.4 & - \\
cold front & - & - & - & - & - & - & 133.2/69 & - & 131/69 & - \\
\hline                                  
\end{tabular}
\end{minipage}
\end{table*}

\begin{table*}
\begin{minipage}[t]{\columnwidth}
\caption{Code performance for different global S/N and binning}            
\label{snresults}
\centering
\renewcommand{\footnoterule}{}
\begin{tabular}{l l l l l l l}       
\hline\hline                
S/N\footnote{Global signal-to-noise ratio} &
$\sigma_{bin}$\footnote{Signal-to-noise ratio per bin} &
$\chi^{2}_{mean}/N_{bins}$\footnote{Using mean densities from 100
  simulation runs} & $\chi^{2}_{errs}/N_{bins}$\footnote{Comparison of mean errors
  with `true' errors -- see text}&$\chi^{2}_{slope}/N_{bins}$\footnote{Using mean logarithmic density
  slope from 100 simulation runs}&$\chi^{2}_{slopeerrs}/N_{bins}$ \\   
\hline                      
15 & 3.0 & 2.2/24 & 37.2/24 & 2.2/23 & 39.2/23 \\  
50 & 3.0 & 1.54/42 & 24.1/42 & 1.3/41 & 25.1/41\\
100 & 3.0 &  1.54/55 & 45.2/55 &2.7/54 & 39.5/54\\
200 & 3.0 & 2.1/71 & 55.5/71 & 2.9/70 & 40.2/70\\
200 & 5.0 & 2.06/45 & 33.5/45 & 1.7/44 & 27.3/44\\
200 & 10.0 & 5.6/33 & 38.4/33 & 3.3/32 & 41.8/32\\
200 & 30.0 & 14.1/17 & 15.1/17 & 9.0/17 & 15.6/17\\
\hline                                  
\end{tabular}
\end{minipage}
\end{table*}

\section{A comparison of {\it XMM-Newton} and {\it Chandra} gas density profiles using the new method}
\label{data}
In the previous section we demonstrated using a range of model density
profiles that our deprojection method performs well in a variety of
situations. We next decided to carry out a comparison of the {\it
Chandra} and {\it XMM-Newton} density profiles obtained using our
method and from analytical models for several clusters for which high
quality data exist from both observatories. The purpose of this
comparison was both to test the performance of our code on real data
and also to test for the first time the consistency of surface
brightness and density profiles obtained from the two observatories.
The comparison of deprojected {\it XMM-Newton} profiles with {\it
Chandra} profiles will be particularly useful as a test of our
PSF-deconvolution method, as the much smaller {\it Chandra} PSF means
that its effects on the central profile are far less important.

We chose to study three nearby clusters Abell 478, Abell 1413 and Abell
1991, which have recently published observations with {\it XMM-Newton}
(Pointecouteau et al. 2004, Pratt \& Arnaud 2002, 2005)
and Chandra (Sun et al. 2003, Vikhlinin et al. 2006). 

\begin{figure}
\begin{center}
\epsfig{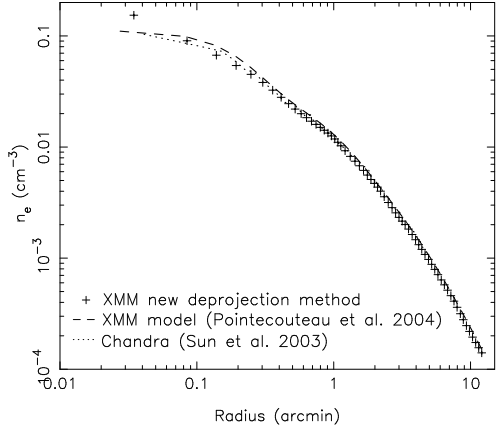}
\caption{Comparison of {\it XMM} and {\it Chandra} profiles for Abell
478 with the {\it XMM} profile obtained with our method (+ symbols,
the best-fitting {\it XMM} model of Pointecouteau et al. (2004)
(dashed line), and the {\it Chandra} profile of Sun et al. (2003)
(dotted line).}
\label{a478}
\end{center}
\end{figure}

For Abell 478, we used the {\it XMM-Newton} surface brightness profile
obtained by Pointecouteau et al. (2004) as input for the deprojection
code. The deprojected {\it XMM-Newton} density profile was then
compared with the analytical model fitted by Pointecouteau et al.
(2004) and with the {\it Chandra} density profile obtained by Sun et
al. (2003). Sun et al. (2003) in fact also used a deprojection method to
obtain their profile; their method uses an ``onion-skin'' technique
without regularisation. Figure~\ref{a478} compares the different
profiles for Abell 478. All three profiles are in good agreement,
although the deprojected {\it XMM} profile is slightly more centrally
peaked than the {\it Chandra} profile.

We used the {\it XMM-Newton} surface brightness profile of Pratt \&
Arnaud (2002) for Abell 1413 as input for the deprojection code. In
this case we compared with the best fitting analytical models of Pratt
\& Arnaud (the `KBB' model) ({\it XMM}) and of Vikhlinin et al. (2006)
({\it Chandra}), which consisted of a $\beta$ model multiplied by a
polytropic model. Figure~\ref{a1413} compares the different density
profiles for Abell 1413. Again, all three density profiles are in good
agreement; however, in this case the model obtained from our
deprojection method is in better agreement with the {\it Chandra}
profiles in the central regions than the {\it XMM} analytical model,
which fails to reproduce the central cuspiness measured by {\it
Chandra}.

\begin{figure}
\begin{center}
\epsfig{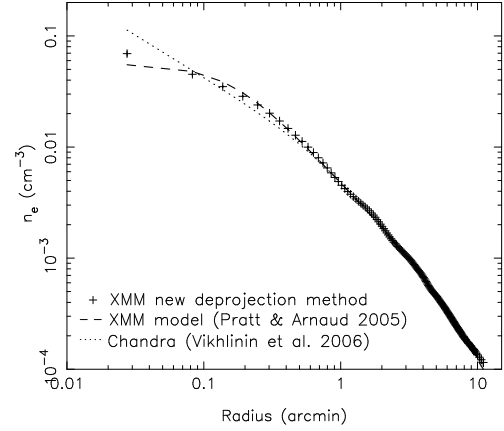}
\caption{Comparison of {\it XMM} and {\it Chandra} profiles for Abell
1991 with the {\it XMM} profile obtained with our method (+ symbols),
the best-fitting {\it XMM} model of Pratt \& Arnaud (2005) (dashed
line), and the {\it Chandra} profile of Vikhlinin et al. (2006)
(dotted line).}
\label{a1991}
\end{center}
\end{figure}

For Abell 1991, we used the {\it XMM-Newton} profile of Pratt \&
Arnaud (2005) as input for the deprojection code, and compared the output
density profile with their analytical model (a sum of two $\beta$
models) and with the best-fitting model to the {\it Chandra} data of
Vikhlinin et al. (2006) (with the same functional form as for Abell
1413, but with an additional second $\beta$-model component). The
results are shown in Figure~\ref{a1991}. In this case, the two
profiles obtained from the {\it XMM-Newton} data are in good
agreement, but both significantly less centrally peaked than the {\it
Chandra} profile. The slope of the {\it Chandra} profile is also
slightly steeper.

\begin{figure}
\begin{center}
\epsfig{figure=a1413_bw.ps,width=6.5cm}
\caption{Comparison of {\it XMM} and {\it Chandra} profiles for Abell
1413 with the {\it XMM} profile obtained with our method (+ symbols),
the best-fitting {\it XMM} model of Pratt \& Arnaud (2002) (dashed line),
and the {\it Chandra} profile of Vikhlinin et al. (2006) (dotted line).}
\label{a1413}
\end{center}
\end{figure}

\begin{figure}
\begin{center}
\epsfig{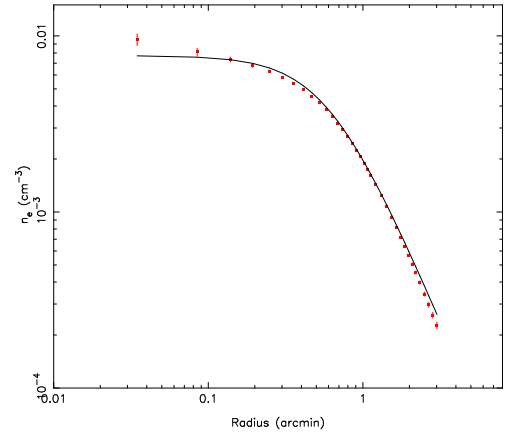}
\caption{Comparison of {\it XMM} profile for CL0016+16 with the {\it
XMM} profile obtained with our method shown in red and the
best-fitting {\it XMM} model of Worrall \& Birkinshaw (2003) in
black.}
\label{cl0016}
\end{center}
\end{figure}

Finally, we also decided to test the code on a more distant cluster
observed by {\it XMM-Newton}, CL0016+16 ($z=0.5455$).
Figure~\ref{cl0016} shows our deprojected density profile with the
best-fitting $\beta$-model profile of Worrall \& Birkinshaw (2003)
both from the {\it XMM-Newton} observation. The profiles are in good
agreement, except at the centre, where Worrall \& Birkinshaw's model
fit had systematic residuals, and in the outermost bins, where our
method identifies a steepening at large radius that could not be taken
into account by the $\beta$-model fit.

\section{A comparison of mass profiles}
\label{dlogne}
\begin{figure*}
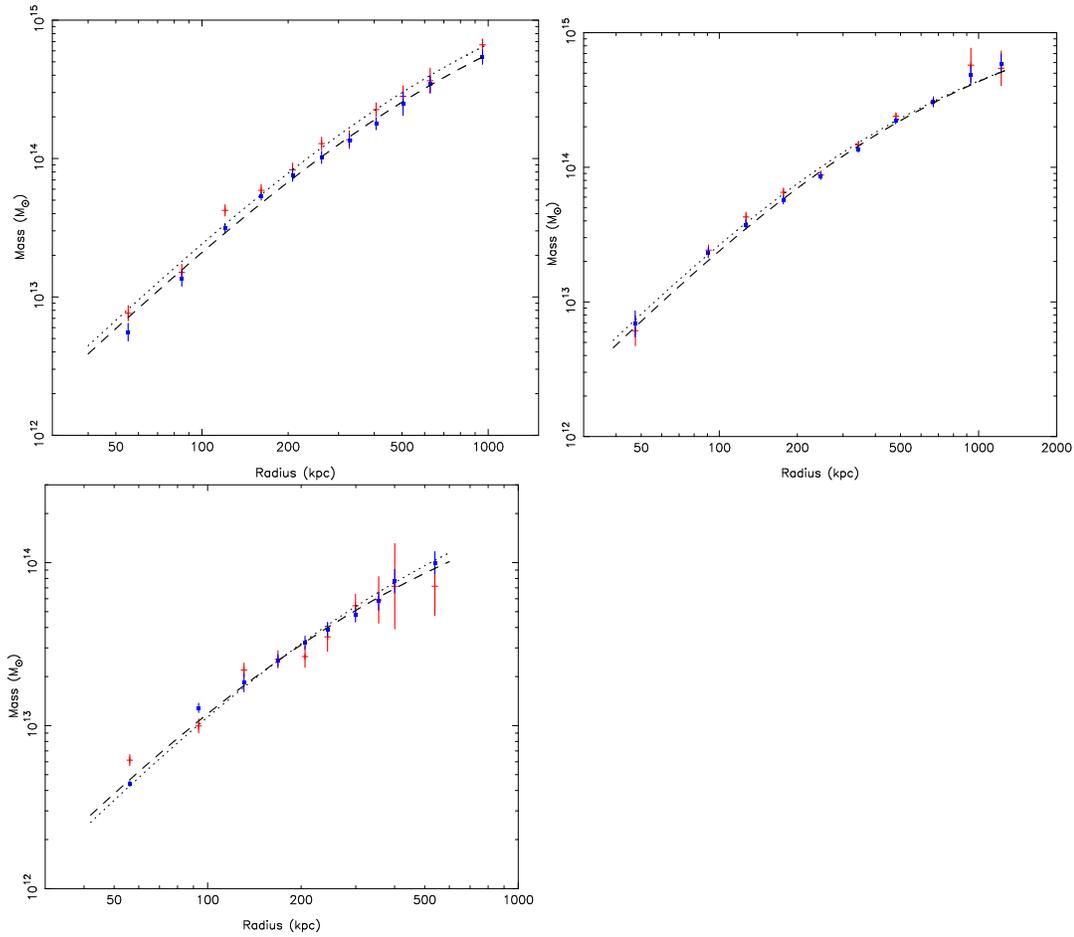

\vbox{
\hbox{
\epsfig{figure=a478_mass.ps,width=7.0cm}
\epsfig{figure=a1413_mass.ps,width=7.0cm}}
\epsfig{figure=a1991_mass.ps,width=7.0cm}
}
\caption{Mass profile comparison for Abell 478 (top left), Abell 1413
  (top right) and Abell 1991 (bottom). In all cases, red + symbols
  indicate the {\it XMM} mass profile obtained by deprojection, blue
  squares the {\it XMM} mass profile obtained by fitting a model to
  the surface brightness distribution (taken from Pointecouteau et al.
  2004, Pratt \& Arnaud 2002, and 2005, respectively). Dashed lines
  indicate NFW fits to the profiles using deprojection, and dotted
  lines best fits to the profiles using analytic density profiles.}
\label{massplots}
\end{figure*}

In addition to comparing the gas density profiles obtained from {\it
  Chandra} and {\it XMM-Newton} cluster observations, as discussed in
  the previous section, we also compared the mass profiles of Abell
  478, Abell 1413 and Abell 1991 obtained from the {\it XMM-Newton}
  observations via the two methods of deprojection and model fitting
  to obtain $d\log(n_e)/d\log(r)$. For both sets of density slope
  profiles, we used the same temperature profiles, those of
  Pointecouteau et al. (2004), Pratt \& Arnaud (2002) and (2005),
  respectively, for Abell 478, Abell 1413 and Abell 1991, to calculate
  a total mass profile. Fig.~\ref{massplots} shows the mass profiles
  for all three clusters. In all three cases the two methods of
  analysis of the {\it XMM-Newton} data obtain similar results. In
  general (and particularly for Abell 1991), the {\it XMM-Newton}
  profiles obtained by model fitting are smoother than those obtained
  via the deprojection method, which is not surprising; however, the
  best-fitting NFW model fits for the new method are consist within
  1$\sigma$ with those reported in the original {\it XMM-Newton}
  analysis papers for each cluster.

\section{Conclusions}

We have described a method for obtaining gas density profiles from
X-ray surface brightness profiles of galaxy clusters, with the aim of
improving constraints on the entropy and total mass distributions of
clusters. We first showed using simulated profiles that this method
performs well in a range of conditions:
\begin{itemize}
\item The effect of the shape of the density profile was tested using
  four models: the $\beta$ model, AB model, KBB model, BB model, as
  well as for profiles with deviations from a smooth shape. We found
  that in all cases the agreement between the output density profile
  and the model was good, with slightly poorer performance in the
  central regions for the KBB and BB models, due to the choice of an
  AB model as a scaling function.
\item The global signal-to-noise of the input profiles does affect the
  accuracy of the output density profiles, with poorer quality data
  giving less accurate results; however, this effect was fairly small,
  and a good agreement ($\chi^{2}$ of 36/24) was obtained for a S/N of
  15, corresponding to a profile containing $\sim 2500$ net counts.
\item The profile binning did not appear to have an important effect
  on the accurate recovery of the input density profile.
\end{itemize}
We then tested the code performance on real {\it XMM-Newton} data for
four clusters: three nearby clusters with published {\it Chandra} gas
density profiles, and one distant cluster. We found that our method
resulted in gas density profiles in better agreement with the higher
resolution {\it Chandra} profiles. We also found that our method
performed well in the case of the distant cluster, CL0016+16,
reproducing a central excess and change of slope at large radii that
could not be taken into account using a $\beta$-model fit. Finally we
demonstrated that the mass profiles obtained from our gas density
profiles are consistent with those obtained from other methods. We
therefore find that our method is suitable for obtaining gas density
profiles both from high signal-to-noise observations of nearby
clusters and also for distant clusters. This model-independent
inversion method will be extremely useful as a consistent and reliable
means of obtaining gas density profiles (and subsequently entropy and
mass profiles) for ongoing studies of large, unbiased samples of
nearby and distant clusters. The deprojection code is available on
request to the authors.

We would like to thank the referee for useful comments.

%

\end{document}